\documentclass[12pt]{article}
\usepackage{geometry}
\usepackage{amsmath}
\usepackage{amssymb}
\usepackage{graphicx}
\usepackage[colorlinks]{hyperref}

\title{Quantum Gravity Phenomenology without Lorentz Invariance Violation: a detailed proposal}
\author{Yuri Bonder\footnote{yuri.bonder@nucleares.unam.mx} and Daniel Sudarsky\footnote{sudarsky@nucleares.unam.mx} 
\\ Instituto de Ciencias Nucleares \\
Universidad Nacional Aut\'{o}noma de
M\'{e}xico
\\ A. Postal 70-543, M\'exico D.F. 04510, M\'exico }
\date{}

\begin{document}

\maketitle

\section*{Abstract}

We describe a scheme for the exploration of quantum gravity phenomenology focussing on effects that could be thought
as arising from a fundamental granularity of space-time. In contrast with the simplest assumptions, such granularity is assumed to respect Lorentz Invariance but is otherwise left unspecified. The proposal is fully observer covariant, it involves
non-trivial couplings of curvature to matter fields and leads to a well defined
phenomenology. We present the effective Hamiltonian which could be used
to analyze concrete experimental situations, some of which are briefly described, and we shortly discuss the degree to which the present proposal is in line with the fundamental ideas behind the equivalence principle.

\section{Introduction}

The idea of accessing empirically some aspects of quantum gravity has, recently, been receiving a lot of attention in connection with possible violations, or deformations, of the space-time symmetries, particularly Lorentz Invariance. The justification for such proposals was essentially the following: It is natural to think that a space-time granularity, presumably associated with the Planck scale, should be incompatible with the Lorentz length contraction of special relativity, and thus, that quantum gravity would lead to either modifications or violations of Lorentz Invariance. This drew the attention of the quantum gravity community, on the general subject of phenomenology of Lorentz Invariance violation, rekindling interest in a program that could be traced to the works \cite{Initial-LIV, SME}. The most direct approach, suggested a real violation of special relativity, associated with the existence of a preferential rest frame (say the one in which the granularity takes
its most symmetric form) and that it is most naturally identified with the local frame singled out by the cosmological co-moving observers (which operationally means the local frame where the CMB dipole vanishes). In this regard, relatively detailed theoretical proposals where made based on String Theory ideas \cite{ST} and others within the Loop Quantum Gravity program \cite{LQG}.
Moreover, a substantial program looking for direct manifestations of these effects has lead to remarkable bounds that essentially rule out the effects which are not suppressed at the level of energies of about a billion times larger than the Planck Scale \cite{Bounds}.
Furthermore, when these ideas are taken at face value and combined with simple quantum field theoretical calculations of the radiative corrections, one finds that the natural size of the effects is not suppressed by the ratio of the characteristic energies to the Planck Energy, as naively expected, but merely by the standard model coupling constants \cite{Collins}. This leads to the conclusion that, if such granular structure of space-time associated with a preferential reference frame were real, the effects would have been noticed long ago. 
One should of course bear in mind that a space-time granularity does not by itself imply a violation of Lorentz Invariance \cite{Sorkin,Rovelli}, and that neither String Theory nor Loop Quantum Gravity
can be said to predict such violations. The only thing that has been argued so far is that they can accommodate them.

Other proposals that are currently popular consider ``deforming" special relativity, so that, while preserving the equivalence of all inertial reference frames,
the notion of ``physical length being equal to the Planck length scale" would remain invariant, something that requires modifying the transformation laws \cite{DSR}. We believe that these proposals face very serious, and perhaps even insurmountable, obstacles \cite{Perspectives}. In view of this situation, one is lead to consider more subtle possibilities where the motivational ideas might be realized without the serious problems faced by the specific proposals mentioned above.

The basic premise which we are interested in addressing in this paper is the following: \textbf{If space-time has some sort of fundamental granularity, which is nonetheless fundamentally respectful of Lorentz symmetry, how could it possibly become manifest?} In a previous work \cite{NewQGP}, a proposal was made in which a granular structure of space-time might become manifest in a rather subtle way so that it would be immune from the previous considerations while still, in principle, susceptible to a phenomenological study. That proposal has a couple of problematic aspects and we deal with them in this article. The basic idea of both, the original and the modified proposals, is based on rather heuristic considerations, which on the other hand, is what one can hope for, given that at this time we do not have anything that resembles a fully satisfactory and successful theory of quantum gravity. 

Regarding notation we use the convention where the space-time metric has signature $(-+++)$. 
Space-time indexes run from $0$ to $3$ and are represented by $\mu$, $\nu$, $\rho$ and $\sigma$; spatial indexes run from $1$ to $3$ and are represented with $i$, $j$, $k$, $l$, $m$ and $n$. The Dirac matrices are
\begin{eqnarray}
\gamma^0=\left(
\begin{array}{cc}
 1 & 0 \\
 0 & -1
\end{array}
\right),\ \ \ \ \ 
\gamma^i=\left(
\begin{array}{cc}
 0 & \sigma_i \\
 -\sigma_i & 0
\end{array}
\right),
\end{eqnarray}
where $\sigma_i$ are the usual Pauli matrices; observe that we use the same convention for the Dirac matrices as Ref. \cite{NRH}. 

The structure of the manuscript is the following: Section \ref{Motivation} is dedicated to motivate the proposal which is given in section \ref{proposal}. The detailed framework where all the calculations are done is described in section \ref{N&C}. One imminent problem of the model and its solution are considered in section \ref{degeneration}. Section \ref{WGC} is dedicated to the study the model in the weak gravity regime and in section \ref{Phen Hamilt} the corresponding phenomenological Hamiltonian is presented. Finally, in sections \ref{ExpOut} and \ref{Conclusion} the experimental outlook of the proposal and the conclusions of the manuscript are presented.

\section{Motivation}\label{Motivation}

Let us start by acknowledging that at this point we have no real good geometrical picture of how a granularity might be associated to space-time while strictly preserving the Lorentz and Poincar\'e symmetries\footnote{The most promising approach in this sense
is the Poset program \cite{Poset} which however is still in its developing stage. It is worth mentioning it because it embodies a
scheme in which the space-time metric is a derived feature and where some objects that are not simply related to it might make
an appearance. That is precisely the kind of situation that our proposal contemplates.} 
and in such situations we cannot rely on anything beyond simple analogies combined with symmetry considerations, which is what we do next.

Consider an experimentalist trying to uncover the granular structure of a salt crystal which he believes has a fundamental symmetry that is anything but cubic, and let us assume that he has chosen to carry out his investigations using a macroscopic crystal that is built to have a cubic symmetry. The researcher is hoping to uncover evidence of the fundamental granularity by looking for experimental signals that indicate deviations from what he considers to be just the macroscopic symmetry of his crystal. We know that he will find none, simply because the fundamental structure of the crystal is also the cubic symmetry. This, we believe, is the situation we face today regarding the fundamental granular structure of space-time (assuming that such granularity exists) and the attempts to seek evidence for its existence through deviations from Lorentz Invariance. As in the case of the crystal, the discrete structure might be studied, of course, but NOT by looking at deviations from the underlying fundamental symmetry. The point is that when one considers a macroscopic crystal whose global form is not compatible with the structure of the fundamental crystal, say spherical, the surface will necessarily include some roughness, and thus, a manifestation of the granular structure will occur through the breakdown of the exact spherical symmetry.

Let us look at these ideas in slightly more detail. It is well known that crystalline minerals can appear in different presentations: In one extreme we have the perfect crystals, where the entire sample is made of units cells arranged in a well defined lattice and in which the symmetry of the primitive unit cell is precisely reproduced by the actual lattice. On the other extreme we have the polycrystals which are made out of a multitude of crystalline grains and where there are many defects and dislocations. We could characterize the interpolating regimes by the number of single grains per unit volume on the sample ($F$). We can similarly define the defect density of the sample ($D$). Furthermore, let us assume that $F$ and $D$ can be combined in a quantity ($G$) that measures the overall local departure from a perfect crystalline structure. Consider now the relationship between the macroscopic shape of a sample and $G$. It is quite clear that in order to have a perfect crystal the macroscopic sample must have a shape that is compatible with that of the underlying crystalline structure. No so for the polycrystalline samples. Thus, the shape of the macroscopic sample clearly sets a lower bound on the $G$ of the sample. In fact, given a very large sample it is possible that $F$ varies from zone to zone in the sample and, if the macroscopic shape varies, the lower limit on $G$ will equally vary along the sample and will be correlated with the local macroscopic geometry. If for some reason the material dynamically evolves while adapting to the macroscopic shape of the sample (maintaining the external shape), there will be an actual correlation between $G$ and the shape, rather than just a correlation between the lower bound of $G$ and the shape of the sample. That is, there would be a correlation between the departure of the geometry of the sample from that dictated by the symmetry of the unit cell and $G$. 

The refined proposal we present in this work, as well as its predecessor, are inspired on the simple picture above. Starting with the assumption that the 
symmetry of the fundamental structure of space-time is itself the Lorentz
symmetry, we would expect no violation of this symmetry at the macroscopic level
to the degree to which the space-time is macroscopically Lorentz Invariant\footnote{We should mention that there exists a point of view that in some sense asks for the existence of some Lorentz Invariant but
nontrivial structure of space-time which would serve as a natural physical mechanism regularizing the divergent integrals in Quantum Field Theory \cite{Collins 2, Joglekar}. 
Such point of view considers that the usual regularization schemes work at the calculational level but that they fail to provide any information about the physics behind their success. In this regard one would seek a regularization scheme that is formulated in a Lorentzian setting (rather than an Euclidean one), involves integer dimensions, and no negative normed states.} on an extended domain. That is, in a region of space-time which could be considered as well approximated by the Minkowski metric, the 
granular structure of the quantum space-time would not become manifest through the breakdown of its symmetry. However, and following our solid state analogy, we are lead to consider the situation in which the macroscopic space-time geometry is not fully matched with the symmetry of its basic constituents.

The main point is that in the event of a failure of the space-time to be exactly Minkowski in
an extended region, the underlying granular structure of quantum gravity could become manifest affecting the propagation of the various matter fields. Such situation should thus be characterized by the Riemann tensor, which is known to describe the failure of a space-time to be Minkowski in an open region. Therefore, the non-vanishing of the Riemann tensor would correspond, at the macroscopic level of description, to the case where the microscopic structure of space-time might become manifest. Moreover, we can expect that due to the implicit correspondence of the macroscopic description with the more fundamental one, that the Riemann tensor would also indicate the space-time directions with which the sought effects would be associated. Furthermore, in contrast with many other proposals of quantum gravity phenomenology, the present one invokes no global preferential frames, no non-dynamical fields, and also implies that there would be no new effects or deviations from standard physics at all in any perfectly Minkowskian region of space-time. 

With these ideas in mind we turn now to the phenomenological proposal. Based on the general considerations above one looks for an effective description of the ways in which the Riemannian curvature could affect, in a nontrivial manner, the propagation of matter fields. The effective description of such a thing should involve Lagrangian terms representing the coupling of matter fields with the Riemann tensor\footnote{It is well known that non-minimal couplings between curvature an matter field arise in the context of quantum field theory in curved space-time, however those effects have a well defined structure \cite{nonminimal} that is quite different from the kind of terms we consider here.}. Before proceeding any further, we 
recall that the Ricci tensor represents that part of the Riemann
tensor which, at least on shell, is locally determined by the
energy momentum of matter fields at the events of interest. Thus, any
coupling of matter fields to the Ricci tensor part of the Riemann tensor
would, at the phenomenological level, reflect a sort of point-wise
self interaction of matter that would amount to a locally defined
renormalization of the usual phenomenological terms, such as a the mass
or the kinetic terms in the Lagrangian. But we are interested in
the underlying structure of space-time rather than the self
interaction of matter. Therefore, we need to ignore the aspects
that encode the latter, which in our case
corresponds to all Lagrangian terms containing the Ricci tensor coupled to
matter fields. The remainder of the Riemann tensor, \textit{i.e.} the Weyl tensor, can thus be thought
to reflect the aspects of the local structure of space-time associated
solely with the gravitational degrees of freedom.

Note that, in the absence of gravitational waves, the Weyl tensor ($W_{\mu\nu\rho\sigma}$) is connected with the nearby ``matter sources" and not just the matter present in the point of interest, and moreover, that such connection involves the propagation of the
influence of such sources through the space-time and thus the structure of the
latter would be playing a central role in the way the influences
might become manifest. In this sense, the Weyl tensor reflects the
``non-local effects" of the matter, in contrast with the Ricci tensor which is fully determined point-wise by the matter fields. The task is then to consider non-trivial ways to couple the Weyl tensor to matter fields in the standard model of particle physics. We will focus here 
on the fermionic fields $\Psi$. The most obvious term one can write is evidently $ W_{\mu\nu\rho\sigma} \bar{\Psi} \gamma^\mu \gamma^\nu \gamma^\rho \gamma^\sigma \Psi$ which unfortunately vanishes 
identically. Therefore we must seek terms that are either highly suppressed\footnote{For instance, the coupling of fermions to the Bel-Robinson tensor or those couplings involving higher derivatives of the fermionic fields are highly suppressed and thus they are phenomenologically
uninteresting.}, or one must consider the possibility of seemingly less natural expressions. Following \cite{NewQGP} we take a particular approach 
of the latter type which, as is shown below, calls for the
eigenvalue analysis of the Weyl tensor together with the use of the volume $4$-form of space-time.

The use of the volume $4$-form in this scheme brings out the question of its naturalness because it involves a choice of orientation that can bring into the scheme features that include the violation of discrete symmetries such as spatial inversion (P) and 
time reversal (T). Our point of view is that the space-time structure should be viewed as involving more than just the metric, and in 
particular it should be considered to include the spatial and time orientations. Namely, that space-time should be regarded not only 
as orientable (so as to permit spin structures, etc.) but as \textit{oriented}, if we want to have a natural explanation for the fact that there exist processes (such as those mediated by the weak interaction) that both, here and on remote regions of space-time, select the same handedness of the fermionic currents
and the same CP (and T) violating phase. In other words, we cannot pretend that space-time has no
information of what is left and right, past and future, and hope to naturally understand that these two notions, as identified by the physics of weak interactions, are the same everywhere. 
Of course, one could choose to view the consistency as being enforced by the fields involved in the
weak interaction themselves, but then one could re-express our view as indicating that it would be inappropriate to think of space-time in the absence of the fields that fill it and probe it \cite{CRS}. In any event, viewing the space and time orientations as intrinsic features of space-time, opens the door to considering tests of whether the quantum gravitational degrees of freedom are sensitive to those features. 

The scheme we propose offers then a natural path for considering violations of discrete symmetries associated with quantum gravity and, although all our experience with gravity seems to point to the absence of such features, it is worth reminding the reader two facts: First, as has been stressed elsewhere \cite{Geometrical}, up to now there have not been direct tests of gravity effects on quantum systems (when considering gravity in the general relativistic context, \textit{i.e.}, as absent when the whole system can be described in a free falling frame, such as in the case
of the COW experiment \cite{COW}). This is a remarkably unpublicized --and often even unrecognized-- feature of our current level of experimental exploration of gravity. Second, the pattern exhibited by all the interactions other than gravity: The weaker the interaction, the less symmetries it respects. Gravity as we know it breaks this pattern quite dramatically. Our proposal contemplates the possibility that at a deeper level this situation would be (at least partially) reversed. 

The actual proposal for the effective interaction is superficially very similar to a part of the Standard Model Extension (SME) of Colladay and Kostelecky \cite{SME} involving what seem like violations of Lorentz Invariance. However, in this scheme, and in contrast with both, the SME and other recent works \cite{NewKostelecky} (which could seem closer in spirit to the present one), the fields are not fixed features of space-time, nor new and independent dynamical fields, but rather dynamical manifestations of gravity determined in each space-time event by the nearby matter distribution (or by a rather distant one in the case of gravity waves). As mentioned before this scheme does not call for global preferential frames, or globally defined non-dynamical tensor fields. The concrete proposal is presented in the following section.
 
\section{A possible non-minimal coupling of Weyl tensor with matter fields} \label{proposal}

Let us consider the type $(2,2)$ Weyl tensor ${W_{\mu\nu}}^{\rho\sigma}$ as a mapping from the space of $2$-forms $\mathcal{S}$ into itself. As is well known,
the space-time metric endows the six dimensional vector space
$\mathcal{S}$ with a pseudo-Riemannian metric of signature $(---+++)$ which we call the supermetric. 
With the aid of the Weyl tensor we can construct two self-adjoint operators on ${\mathcal{S}}$ 
\begin{eqnarray}
{{(W_+)}_{\mu\nu}}^{\rho\sigma}&\equiv&\frac{1}{2}\left({W_{\mu\nu}}^{\rho\sigma}+{W^\dagger_{\mu\nu}}^{\rho\sigma}\right),\\
{{(W_-)}_{\mu\nu}}^{\rho\sigma}&\equiv&\frac{i}{2}\left({W_{\mu\nu}}^{\rho\sigma}-{W^\dagger_{\mu\nu}}^{\rho\sigma}\right),
\end{eqnarray}
where ${W^\dagger_{\mu\nu}}^{\rho\sigma}$ represents the adjoint of ${W_{\mu\nu}}^{\rho\sigma}$. This operators can be diagonalized having a complete set of eigenforms. The non-null eigenforms of the tensors ${{(W_\pm)}_{\mu\nu}}^{\rho\sigma}$,
$\Xi_{\mu \nu}^{(\pm,s)}$, corresponding to non-vanishing eigenvalues
$\lambda^{(\pm,s)}$, once properly normalized to $\pm1$ in terms of the supermetric, can be used to construct the types of Lagrangian terms we are
interested in. The sign in the superindex $(\pm,s)$ denotes which tensor we are considering while its different eigenvalues and eigenforms are labeled with $s=1,\ldots,6$. In order to proceed, we recall that the eigenvalues $\lambda^{(\pm,s)}$ have the dimension of the Weyl tensor so that ${\lambda^{(\pm,s)}}^{1/2}/M_{\rm Pl}$ is dimensionless, where $M_{\rm Pl}$ represents Planck's mass. We use $r^{(\pm,s)}$ to indicate the power at which $M_{\rm Pl}$ enters at the phenomenological level. That is we multiply the terms with the correct dimensions that can appear in the Lagrangian by $({\lambda^{(\pm,s)}}^{1/2}/M_{\rm Pl})^{r^{(\pm,s)}}$. The appearance of the square root of the eigenvalues might seem very unnatural but we must recall that we are taking the view that the metric (and therefore the Weyl tensor) is far from the fundamental object underlying the true gravitational degrees of freedom. This is in fact the view taken for instance in the Loop Quantum Gravity approach and it is even more accentuated in approaches like the area-metric theories \cite{Schuller}. On the same token, the expectation that such terms must always be suppressed by at least one or two factors of $M_{\rm Pl}$ has been disproved in the analysis of \cite{Collins}. For massive particles we have several other ratios involving the particle's mass at our disposal but given the fact that they can always be included as part of the dimensionless coupling constants, we do not insert any such factors explicitly. However, we do have to take into account a possible flavor dependence, which could arise not only from the different masses but also from the detailed way the different fields interact with the virtual excitations that intimately probe the underlying space-time structure. The most general form for the coupling term of the $\lambda^{(\pm,s)}$ and $\Xi_{\mu \nu}^{(\pm,s)}$ with fermions in these schema\footnote{The original proposal of \cite{NewQGP} did not involve separation of the Weyl tensor into its self adjoint components and lead in principle to potentially non hermitian terms in the Hamiltonian. This does not occur in the present refinement.} is:
\begin{eqnarray} \label{Lagrangian0}
{\cal L}=\sum_{a} \sum_{s=1}^6 \sum_{\alpha=\pm} \left[ \xi^{(\alpha)}_a{\lambda^{(\alpha,s)}}^{1/2} \left(\frac{{\lambda^{(\alpha,s)}}^{1/2}}{M_{\rm Pl}} \right)^{r^{(\alpha,s)}} \Xi^{(\alpha,s)}_{\mu \nu}\right] \bar\Psi_a \gamma^{\mu}\gamma^{\nu}\Psi_a,
\end{eqnarray}
where $a$ denotes flavor, $\xi^{(\pm)}_a$ are some dimensionless coupling constants and $r^{(\pm,s)}$ are constants required to be greater than $-1$ to ensure that in the limit of flat space-time the new terms vanish. 

One important feature of this proposal is that, ignoring the degrees of freedom of space-time itself, these Lagrangian terms have dimension 3 and are thus super-renormalizable. Therefore the radiative corrections associated only with the fields of the standard model of particle physics (\textit{i.e.}, disregarding gravity which needless to say has yet to be successfully quantized) do not generate large problematic corrections. In other words, if we use this new interaction as part of the radiative correction calculations involving only the standard model of particle physics and no quantum gravitational corrections, which have problems of their own, we are not lead to large or unsuppressed effects.
This is a very important feature because, as we already mentioned, such radiative corrections were what lead to very serious problems for the phenomenological proposals involving Lorentz Invariance violations. 

\section{The detailed framework} \label{N&C}

We use local Minkowskian coordinates around a space-time event $p$ where we wish to probe for the effects (we are actually thinking of Riemannian normal coordinates about that point). We introduce a tensorial notation for tensors over $\mathcal{S}$ where capital letters $A$, $B$, $C$ and $D$ represent antisymmetric pairs of space-time indexes which are numerated with roman numerals with the convention $I=01$, $II=02$, $III=03$, $IV=23$, $V=31$ and $VI=12$. These new indexes are ``lowered" by the supermetric
\begin{equation} \label{supermetric}
\mathcal{G}_{AB}=\mathcal{G}_{\mu\nu\rho\sigma} \equiv (g_{\mu\rho} g_{\nu\sigma}-g_{\mu\sigma} g_{\nu\rho}),
\end{equation}
and ``raised" by $\mathcal{G}^{AB}=\mathcal{G}^{\mu\nu\rho\sigma}$. Note that the contraction of capital-letter indexes differs from the contraction of the corresponding space-time indexes by a factor $1/2$. This discrepancy is ignored since this factor can be absorbed by rescaling the coupling parameters.

Any $(2,2)$ tensor ${T_{\mu\nu}}^{\rho\sigma}$ which is antisymmetric in its two pairs of indexes can be expressed as a $6 \times 6$ matrix given by
\begin{eqnarray}
{T_A}^B \equiv \left(\begin{array}{cccc}
{T_I}^I & {T_I}^{II} &\cdots & {T_I}^{VI} \\
{T_{II}}^I & {T_{II}}^{II} & \cdots & {T_{II}}^{VI} \\
& & & \\
\vdots & \vdots & \ddots & \vdots \\
& & & \\
{T_{VI}}^I & {T_{VI}}^{II} & \cdots & {T_{VI}}^{VI} 
\end{array} \right)= \left(\begin{array}{cccccc}
&&&&&\\
&\mathbf{K}&&&\mathbf{L}&\\
&&&&&\\
&&&&&\\
&\mathbf{M}&&&\mathbf{N}&\\
&&&&&
\end{array} \right),
\end{eqnarray}
where boldface capital letters represent $3\times 3$ matrices. In particular, the supermetric at the event $p$ has the form
\begin{eqnarray}
\mathcal{G}_{AB}=\left(\begin{array}{cccccc}
&&&&&\\
&-\mathbf{ 1}&&&\mathbf{ 0}&\\
&&&&&\\
&&&&&\\
&\mathbf{0}&&&\mathbf{1}&\\
&&&&&
\end{array} \right)=\mathcal{G}^{AB}
\end{eqnarray}
where $\mathbf{ 0}$ and $\mathbf{1}$ respectively stand for the $3 \times 3$ zero and identity matrices. 
In addition, the natural volume element $\epsilon_{\mu\nu\rho\sigma}$ associated with the space-time metric can be expressed locally as
\begin{eqnarray} \label{epsil}
{\epsilon_A}^B = \left(\begin{array}{cccccc}
&&&&&\\
&\mathbf{ 0}&&&\mathbf{ 1}&\\
&&&&&\\
&&&&&\\
&-\mathbf{1}&&&\mathbf{0}&\\
&&&&&
\end{array} \right)=-{(\epsilon^{-1})_A}^B .
\end{eqnarray}
Moreover, $2$-forms in the capital-letters index notation can be expressed using $3$-component column vectors (marked with an arrow) as
\begin{eqnarray}
X_A=\left(\begin{array}{c} X_I\\ \vdots \\ X_{VI}\end{array} \right)=\left(\begin{array}{c} \vec{u}\\ \\ \vec{v} \end{array} \right).
\end{eqnarray}

Due to the its symmetries and the traceless property, the Weyl tensor expressed as a $6 \times 6$ matrix has, in these coordinates, the generic form \cite{Weyl matrix form} 
\begin{eqnarray} \label{Weylgral}
{W_A}^B= \left(\begin{array}{cccccc}
&&&&&\\
&\mathbf{ A}&&&\mathbf{ B}&\\
&&&&&\\
&&&&&\\
&-\mathbf{B}&&&\mathbf{ A}&\\
&&&&&
\end{array} \right),
\end{eqnarray}
where $\mathbf{ A}$ and $\mathbf{ B}$ are $3 \times 3$ real traceless symmetric matrices. Therefore,
 the hermitian operators have the matricial forms:
\begin{eqnarray} 
{{(W_+)}_{A}}^{B}= \left(\begin{array}{cccccc}
&&&&&\\
&\mathbf{ A}&&&\mathbf{ 0}&\\
&&&&&\\
&&&&&\\
&\mathbf{0}&&&\mathbf{ A}&\\
&&&&&
\end{array} \right),
\end{eqnarray}
and 
\begin{eqnarray} 
{{(W_-)}_{A}}^{B}= \left(\begin{array}{cccccc}
&&&&&\\
&\mathbf{ 0}&&&i\mathbf{ B}&\\
&&&&&\\
&&&&&\\
&-i\mathbf{B}&&&\mathbf{ 0}&\\
&&&&&
\end{array} \right).
\end{eqnarray}
Finally, we expand the matrices $\mathbf{A}$ and $\mathbf{B}$ in powers of $1/c$ as
\begin{eqnarray}
\label{expA} \mathbf{A}&=&\mathbf{A}^{[0]}+\frac{1}{c}\mathbf{A}^{[1]}+\cdots,\\
\label{expB} \mathbf{B}&=&\mathbf{B}^{[0]}+\frac{1}{c}\mathbf{B}^{[1]}+\cdots,
\end{eqnarray}
and work in the lowest order of this expansion.

\section{Weyl tensor degeneration} \label{degeneration}

It must be noted that the construction of the phenomenological Lagrangian presented in equation (\ref {Lagrangian0}) is well defined only as long as there are no degeneracies in the ${{(W_\pm)}_{A}}^{B}$ tensors. In fact, if there are two eigenforms $ \Xi^{(\pm,1)}_{\mu \nu}$ and $ \Xi^{(\pm,2)}_{\mu \nu}$ with the same eigenvalue $\lambda^{(\pm)}$, then so is any linear combination $\alpha \Xi^{(\pm,1)}_{\mu \nu}+ \beta\Xi^{(\pm,2)}_{\mu \nu}$ where the coefficients $\alpha$ and $\beta$ are fixed to satisfy the appropriate normalization conditions. The key issue is that, at this point, one would not know which of this linear combinations the model calls for. In the original proposal \cite{NewQGP}, this aspect was noted (thanks to a comment from a referee) and then it was assumed that such degeneracies would be rather unusual and the scheme was supposed to refer only to the non-degenerate case. However, even in this case there is always a structural degeneration which can be understood as a result of the identities
\begin{eqnarray}
{\epsilon_A}^B {W_B}^C {{(\epsilon^{-1})}_C}^D&=& {W_A}^D,\\
{\epsilon_A}^B {W^\dagger_B}^C {{(\epsilon^{-1})}_C}^D&=& {W^\dagger_A}^D,
\end{eqnarray}
which imply that
\begin{equation}\label{Weyl+-degeneration}
{\epsilon_A}^B {{(W_\pm)}_B}^C {{(\epsilon^{-1})}_C}^D= {{(W_\pm)}_A}^D.
\end{equation}
The point is that if $\Xi^{(\pm)}_A$ is an eigenform of ${{(W_\pm)}_B}^C$ with a given eigenvalue, $ {\epsilon_A}^B \Xi^{(\pm)}_B$ is a degenerated eigenform of ${{(W_\pm)}_B}^C$. We work under the assumption that there are no further degeneracies, \textit{i.e.}, we only consider type I space-times according to the Petrov classification where the matrices $\mathbf{A}$ and $\mathbf{B}$ are non-degenerated \cite{Weyl matrix form}. Clearly in this particular case each of the hermitian operators ${{(W_\pm)}_B}^C$ has three\footnote{The traceless property of these matrices implies that the eigenvalues are not independent but this has no impact on the degeneracy which is our only concern here.} different (real) eigenvalues which are labeled by $\lambda^{(\pm,l)}$ where the sign indicates the corresponding operator and $l=1,2,3$.

It is noteworthy that we can make use of the object ${\epsilon_A}^B$ which leads to this degeneration 
to provide for a refinement of the recipe that eliminates this problem. We select within each degeneracy subspace the eigenforms satisfying the requirement
\begin{equation} \label{eZZ}
\epsilon^{AB} \Xi_{A}^{(\pm,l)}\Xi_{B}^{(\pm,l)}=0.
\end{equation}
The condition above selects two rays 
in each degeneracy subspace, one of the rays contains non-negative normed vectors and the other contains only non-positive normed ones. Within each ray we choose the representatives satisfying equation (\ref{eZZ}) and such that
\begin{equation} \label{eZZN}
\mathcal{G}^{AB}\Xi_{A}^{(\pm,l)}\Xi_{B}^{(\pm,l)}=-1,
\end{equation} 
and then define the positive norm eigenform (an identify it with the tilde) corresponding to the same eigenvalue as
\begin{equation} \label{eZZP}
\widetilde\Xi_{A}^{(\pm,l)} \equiv {\epsilon_A}^B\Xi_{B}^{(\pm,l)}, 
\end{equation}
that automatically has norm $+1$.
 
The refined proposal then calls precisely for the use of these particular forms in the construction of the Lagrangian (\ref {Lagrangian0}), thus removing the problematic arbitrariness identified at the beginning of this section. 
 
There is still one missing aspect that we would need to fix in order to have a truly unambiguous recipe for writing the desired Lagrangian term: A sign ambiguity for each term.
That is, if $\Xi_A^{(\pm,l)}$ is an eigenform of ${{(W_{\pm})}_A}^B$ satisfying the conditions (\ref{eZZ}) and (\ref{eZZN}), then $-\Xi_A^{(\pm,l)}$ is another eigenform satisfying the same conditions. In other words, we need to fix the signs of the eigenforms of ${{(W_{\pm})}_A}^B$ using the geometrical structure at hand: The volume four-form, the time orientation, the metric and Riemann tensors, and perhaps some higher derivatives thereof. We have not been able to find any reasonably simple way to do this that can be called canonical, which is what we are looking for. We must acknowledge at this point that it is possible that no such recipe will lead to a continuous dependence of these signs on the input quantities (topological stability), or even that no such canonical prescription exists at all. In such event we could still hope that something like a ``spontaneous symmetry breaking'' mechanism might be at the source of the selection of the underlying space-time microstructure behind the macroscopic manifestation that we describe in terms of ``the metric", and in that situation the macroscopic structures would not be enough to determine every detail of the way the microstructure could manifest phenomenologically. In particular, in our case and point, the macrostructure would not determine the signs in question.
Fortunately, the fact that the remaining problem is related only to these overall signs allows in practice for a simple solution: The introduction of \textit{a priori} distinct coupling constants, whose signs are undefined. Thus we re-write equation (\ref{Lagrangian0}) as
\begin{eqnarray} \label{Lagrangian}
\mathcal{L}=\sum_{a} \sum_{l=1}^3\sum_{\alpha = \pm}
\left[ \xi_{a}^{(\alpha,l)} {\lambda^{(\alpha,l)}}^{1/2} \left(\frac{{\lambda^{(\alpha,l)}}^{1/2}}{M_{\rm Pl}} \right)^{r^{(\alpha,l)} }\Xi^{(\alpha, l)}_{\mu \nu}\right.\nonumber\\\left.
+\widetilde{\xi}_{a}^{(\alpha,l)} {\lambda^{(\alpha,l)}}^{1/2}\left(\frac{{\lambda^{(\alpha,l)}}^{1/2}}{M_{\rm Pl}} \right)^{\widetilde{r}^{(\alpha,l)}}\widetilde{\Xi}^{(\alpha,l)}_{\mu \nu}
\right] \bar\Psi_a \gamma^{\mu}\gamma^{\nu}\Psi_a,
\end{eqnarray}
where the index $a$ labels the fermion flavor, $\alpha$ runs over $\pm$ and $\xi_a^{(\pm,l)}$, $\widetilde{\xi}_a^{(\pm,l)}$, $r^{(\pm,l)}$ and $\widetilde{r}^{(\pm,l)}$ are the free parameters of the model. It should be noted that this interaction Lagrangian is fully covariant, and does not involve any non-dynamical fields that could be thought as defining a locally preferred frame or space-time direction.

\section{The Weak Gravity Regime}\label{WGC}

We consider here the situations in which the linearized gravity approximation is justified as it will be in all conceivable experiments in the solar system and particularly in a laboratory. We are interested on the effects of the gravitational environment on test particles and therefore, when describing how the local matter distribution determines the gravitational environment, we only need to consider the standard Einstein's equation\footnote{This is justified because on the one hand the 
 corrections on Einstein's equations that would arise from the new couplings are of higher order in the gravitational constant $G$, but even more importantly because the ordinary bulk matter in a laboratory is not polarized.}. In this regime, it is enough to consider the lowest order perturbative analysis, so we write 
\begin{equation}
g_{\mu\nu}=\eta_{\mu\nu}+ \gamma_{\mu\nu}, 
\end{equation}
where $\eta_{\mu\nu}$ is a flat space-time metric and $\gamma_{\mu\nu}$ is a small perturbation. We take the Minkowskian coordinates associated with $\eta_{\mu\nu}$ as approximately identified with the laboratory measured coordinates: $t, \vec x $. After fixing the gauge in the standard fashion, namely by imposing $\partial^\mu \bar{\gamma}_{\mu\nu}=0$ where $\bar{\gamma}_{\mu\nu}\equiv\gamma_{\mu\nu}-\frac{1}{2}\eta_{\mu\nu}\eta^{\rho\sigma}\gamma_{\rho\sigma}$ (see \cite{Wald}), we focus on the structure of the Weyl tensor. As indicated, we only consider the situations in which the sources vanish at the points to be probed experimentally, thus $T_{\mu\nu}=0$ and $W_{\mu\nu\rho\sigma}=R_{\mu\nu\rho\sigma}$. Therefore, the Weyl tensor can be expressed as
\begin{equation} \label{weyllin2} 
{W_{\mu\nu}}^{\rho\sigma}=-2\partial^{[\rho} \partial_{[\mu} \bar{\gamma}_{\nu]}^{\sigma]}+\delta^{[\rho}_{[\mu} \partial_{\nu]} \partial^{\sigma]} \bar{\gamma},
\end{equation}
where $\bar{\gamma} \equiv\eta^{\mu\nu}\bar{\gamma}_{\mu\nu}$. The linearized Einstein equations are 
\begin{equation} \label{Helmholtz}
\partial_\rho \partial^\rho \bar{\gamma}_{\mu\nu}=-16\pi G T_{\mu\nu},
\end{equation}
where $G$ stands for the gravitational constant. The appropriate solution is obtained in the standard way by means of the retarded Green's function:
\begin{equation} \label{Green}
\bar{ \gamma}_{\mu\nu}=4G\int T_{\mu\nu}(x')\frac{\delta(t'-t_r)}{R}d^4x',
\end{equation}
where $R \equiv |\vec{x}-\vec{x}'|$ and $t_r \equiv t- R/c$ is the retarded time.
To lowest order in $1/c$, we write the components of the energy momentum tensor as $T_{\mu\nu}=\delta_\mu^0 \delta_\nu^0 \rho$ where $\rho$ is the matter density. From equation (\ref{Green}) it is easy to see that $\bar{ \gamma}_{\mu\nu}=\delta_\mu^0 \delta_\nu^0 \bar{ \gamma}_{00}$, thus, the Weyl tensor is re-written as
\begin{equation} \label{weyllin4}
{W_{\mu\nu}}^{\rho\sigma}=\left(2 \delta_{[\mu}^{0}\delta_{\nu]}^{i} \delta^{[\rho}_{0} \delta^{\sigma]}_j - \delta_{[\mu}^{[\rho} \delta_{\nu]}^{|i|}\delta^{\sigma]}_j \right)\partial_i \partial^j \bar{ \gamma}_{00}+\mathcal{O}\left(\frac{1}{c}\right).
\end{equation}
In addition, it is easy to see from equation (\ref{Green}) that in this regime
\begin{equation}\label{gamma00}
\bar{ \gamma}_{00}=4G \int \frac{\rho(\vec{x}', t)}{R}d^3x'= 4\Phi_N,
\end{equation}
where the $\Phi_N$ is the ordinary Newtonian potential due to the matter source. We define 
\begin{equation}
{Q_{i}}^j\equiv \partial_i \partial^j\Phi_N -\frac{1}{3} \delta_i^j \partial_k \partial^k \Phi_N
\label{tidal}
\end{equation}
which is a traceless symmetric tensor. Note however that, since we are considering the field due to matter other than the one present in the point of interest (\textit{i.e.}, the point where the probe is), as we argued when we said we should focus on the Weyl tensor, the last term in this expression, which is proportional to the density of matter at the point where the probe is, does not contribute to the calculation. Substituting equation (\ref{gamma00}) in equation (\ref{weyllin4}) we find:
\begin{eqnarray}
\label{compWeyllin1} {W_{0l}}^{0n}&=& {Q_l}^n+\mathcal{O}\left(\frac{1}{c}\right),\\ 
\label{compWeyllin2}{W_{kl}}^{mn}&=&-4\delta_{[k}^{[m}{Q_{l]}}^{n]}+\mathcal{O}\left(\frac{1}{c}\right),\\
\label{compWeyllin3}{W_{0l}}^{mn}&=& \mathcal{O}\left(\frac{1}{c}\right),\\
\label{compWeyllin4}{W_{kl}}^{0n}&=& \mathcal{O}\left(\frac{1}{c}\right).
\end{eqnarray}
Comparing with equation (\ref{Weylgral}) we get 
\begin{eqnarray} \label{A}
{{A^{[0]}}_j}^k&=& {Q_j}^k,\\
{{B^{[0]}}_j}^k &=&0,
\end{eqnarray}
where the expansions (\ref{expA}) and (\ref{expB}) are used. Therefore, ${{(W_-)}_A}^B=\mathcal{O}(1/c)$ and at zero-th order in $1/c$ we can neglect the contribution of its eigenvalues and eigenforms. Furthermore, the eigenvalues and eigenforms of ${{(W_+)}_A}^B$ are obtained when diagonalizing the matrix $\mathbf{A}^{[0]}$ (or equivalently, the matrix whose components are ${Q_j}^k$.) That is, we need to find $\alpha^{(l)}$ and ${a^{(l)}}^j$ such that
\begin{equation}
\label{eigena} {Q_j}^k {a^{(l)}}^j=\alpha^{(l)}{a^{(l)}}^k.
\end{equation}
Then, $\lambda^{(+,l)}=\alpha^{(l)}$
and
\begin{eqnarray}
\Xi^{(+,l)}_{A}= \left(\begin{array}{c}\vec{a}^{(l)} \\ \\ \vec{0}\end{array} \right),
\end{eqnarray}
where the arrow represents $3$-component column vectors. Note that the hermiticity of matrix $\mathbf{A}$ (and hence of $\mathbf{A}^{[0]}$) together with the normalization condition leads to $\delta_{jk}{a^{(l)}}^j{a^{(l)}}^k=\delta^{lm}$. Also note that the condition (\ref{eZZ}) is automatically satisfied by the given eigenforms. The tilded (or positive normed) eigenforms for the corresponding eigenvalues are then 
\begin{eqnarray}
\widetilde{\Xi}^{(+,l)}_{A}= \left(\begin{array}{c}\vec{0} \\ \\ \vec{a}^{(l)} \end{array} \right).
\end{eqnarray}

\section{The phenomenological Hamiltonian}\label{Phen Hamilt}

In order to obtain a Hamiltonian which can be used in the analysis of experimental tests we note that, in principle, we have
the same types of effects considered in the
SME \cite{SME} containing just Lagrangian terms
the form $-1/2 H_{\mu \nu} \bar\Psi \sigma^{\mu \nu} \Psi$,
where $\sigma^{\mu \nu}=[\gamma^\mu, \gamma^\nu ]/2$. Comparing with equation (\ref{Lagrangian}), we can connect the two formalisms through the local identification:
\begin{eqnarray} \label{Hmunu}
H_{\mu \nu}=-2 \sum_a\sum_{l=1}^3 {\lambda^{(+,l)}}^{1/2} \left\{ \xi_{a}^{(+,l)} \left(\frac{{\lambda^{(+,l)}}^{1/2}}{M_{\rm Pl}} \right)^{r^{(+,l)}} \Xi^{(+,l)}_{\mu \nu}\right.\nonumber\\\left.+\widetilde{\xi}_{a}^{(+,l)} \left(\frac{{\lambda^{(+,l)}}^{1/2}}{M_{\rm Pl}} \right)^{\widetilde{r}^{(+,l)}} \widetilde{\Xi}^{(+,l)}_{\mu \nu} \right\}+\mathcal{O}\left(\frac{1}{c}\right).
\end{eqnarray}
It should be stressed however, that in contrast with the SME scheme, our proposal leads to a predetermined space-time dependence of the effects which is dictated by the
surrounding gravitational environment. For simplicity we restrict to the case where all the test particles are electrons. This way there are no sums over the fermion flavor and the index $a$ can be omitted.

The relevant correction in the non-relativistic Hamiltonian for electrons can be directly read of from equation (\ref{Lagrangian})
using the formulation of \cite{NRH} as
\begin{equation} \label{HNR}
{\cal H}_{NR}= \epsilon^{ijk} \left[\frac{1}{2}\left(\sigma_i + \left(\vec{\sigma}\cdot \frac{\vec{P}}{m} \right) \frac{P_i}{m} \right) H_{jk} + \left(1- \frac{1}{2}\frac{P^2}{m^2} \right) \frac{P_i}{m} \sigma_j H_{0k} \right],
\end{equation}
where
$\vec{P}$ and $m$ are respectively the momentum and mass of the test particle and the $\sigma_i$ stand for the Pauli matrices. Note that it would be natural to expect that the term proportional to $H_{0k}$
could only arise from a non time reversal symmetric aspect of the source which in the case of ground based experimental setups would seem to entail the Earth's rotation, but in fact in this scheme, due to the role that could be played by the time and space orientations, as discussed above, this is not the case and such terms could arise even in the absence of explicit time orientation in the inducing sources. 

To write the perturbation Hamiltonian of equation (\ref{HNR}) in terms of the eigenvalues and eigenforms of the Weyl tensor it is useful to define the $3$-vectors
\begin{eqnarray}
D^i &\equiv& \frac{1}{2} \epsilon^{ijk} H_{jk},\\
F^i &\equiv& {H_0}^i,
\end{eqnarray}
where $H_{\mu\nu}$ is given by equation (\ref{Hmunu}). When the substitution is done and after some rearrangement, the vectors $D^i$ and $F^i$ are respectively
\begin{eqnarray} 
\label{vecD} \vec{D}&=&-2\sum_{l=1}^3{\lambda^{(+,l)}}^{1/2} \widetilde{\xi}^{(+,l)} \left(\frac{{\lambda^{(+,l)}}^{1/2}}{M_{\rm Pl}} \right)^{\widetilde{r}^{(+,l)}} \vec{a}^{(l)}+\mathcal{O}\left(\frac{1}{c}\right) , \\
\label{vecF} \vec{F}&=&-2\sum_{l=1}^3 {\lambda^{(+,l)}}^{1/2} \xi^{(+,l)} \left(\frac{{\lambda^{(+,l)}}^{1/2}}{M_{\rm Pl}} \right)^{r^{(+,l)}} \vec{a}^{(l)}+\mathcal{O}\left(\frac{1}{c}\right), 
\end{eqnarray}
where $\lambda^{(+,l)}$ and $ \vec{a}^{(l)}$ are the eigenvalues and normalized eigenforms of the gravity tidal $3\times 3$ matrix of equation (\ref{tidal}).

Then, the Hamiltonian term given in equation (\ref{HNR}) can be expressed as,
\begin{equation} \label{HNRfinal}
{\cal H}_{NR}= \vec{\sigma} \cdot \vec{D} + \left(\vec{\sigma}\cdot \frac{\vec{P}}{m} \right) \left(\frac{\vec{P}}{m} \cdot \vec{D}\right)+ \left(1- \frac{1}{2}\frac{P^2}{m^2} \right) \frac{\vec{P}}{m} \cdot \vec{\sigma} \times \vec{F} ,
\end{equation}
where
$\vec{P}$ and $m$ are respectively the momentum and mass of the test particle and $\vec{\sigma}$ is the vector formed with Pauli matrices. In addition, the standard (euclidian) interior and exterior product of $3$-vectors is represented with $\cdot$ and $\times$, respectively. Expression (\ref{HNRfinal}) can now be directly compared with the outcomes of concrete experiments in order to obtain bounds for $\xi_a^{(+,l)}$, $\widetilde{\xi}_a^{(+,l)}$, $r^{(+,l)}$ and $\widetilde{r}^{(+,l)}$, or more optimistically, to look for a quantum gravity signal.

\section{Experimental outlook}\label{ExpOut}

It is clear that the relevant experiments must be associated with both, relative large gravitational tidal effects in the local environment (indicating large curvature) together with probes involving polarized matter as the explicit appearance of the Dirac matrix $[\gamma^\mu, \gamma^{\nu}]$ indicates. Both conditions seem from the onset difficult to achieve and to control. Polarized matter is usually highly magnetic and thus electromagnetic disturbances would need to be controlled to a very high degree as they would tend to obscure any possible effects. In addition, gravitational field gradients are usually exceedingly small on Earth. Moreover, the fact that these gradients might vary sharply from location to location on Earth indicates that extreme care must be taken when comparing one experiment to another carried out under what seem to be slightly different circumstances and environments\footnote{This is a reminiscent of the situation encountered with the studies of the ``Fifth Force" proposals \cite{Ephraim}.}.

Let us first discuss the kind of experiments that cannot detect the proposed effects. These are the so call Hughes-Drever experiments \cite{HD} where a set of nuclear Zeeman transition lines are monitored during several days leading to sharp bounds on the presence of anisotropic sidereal features having a 24 hour period. Notice that the gravitational gradients are dominated by the local matter distribution which does not have such sidereal dependence and therefore the effects due to the coupling with Weyl cannot be detected in this kind of experiment. However, one could imagine setting up a similar experiment where a sizable mass, arranged so as to produce relatively large gravitational gradients is made to move around the experimental setup with a given period and one would look for a signal with the chosen frequency. These sort of experimental setups naturally involve the risk that a signal with the desired periodicity but resulting from a undesirable ordinary coupling would be picked up in the apparatus. Such problems can be addressed by suitable geometrical arrangements where the signal of interest would have a frequency that is a nontrivial multiple of the actual frequency of mechanical motions and electrical signals involved, in analogy with the technique used in modern E\"otWash experiments \cite{EoTWash}.

One very interesting possibility for a relevant experiment is provided by the construction of probes that have large spin polarization with extremely low magnetization as those done in the group led by E.G. Adelberger \cite{Adelberger}. Their torsion balance, which has been built to avoid the effects of gravity gradients, would need to be modified but this seems to be a rather feasible possibility. 
Without getting into the experimental details one can consider the magnitude of the effect as a function of the phenomenological parameters simply by evaluating the energy difference associated with the two orientations of the spin of an electron in a laboratory induced gravitational tidal field, and comparing it with the energy difference for the two spin orientations in the magnetic field of the Earth. 
We now consider a possible experimental setup using the tidal field generated by two spherical bodies of mass $m$ whose centers are located horizontally at a distance $d$ from the probe. Observe that we can treat the source masses and the Earth as point-like particles since we are only interested in the tidal effects at the location of the probe. Using an euclidian reference frame where the probe is at the origin and the sources and the center of the Earth are on the $x$ and $z$ axis, respectively, the Newtonian potential $\Phi_N$ can be easily computed. Moreover, with the aid of equation (\ref{tidal}) we find 
\begin{eqnarray}
\mathbf{A}^{[0]}=G\left(
\begin{array}{ccc}
  \frac{4 m}{d^3}-\frac{M}{D^3} & 0 & 0 \\
 0 & -\frac{2 m}{d^3}-\frac{M}{D^3} & 0 \\
 0 & 0 & \frac{2 M}{D^3}-\frac{2 m}{d^3}
\end{array}
\right)
\end{eqnarray}
where $M$ and $D$ are the mass and radius of the Earth, respectively. Note that $\mathbf{A}^{[0]}$ is traceless and non-degenerated, as expected. Choosing the phenomenological parameters $m$ and $d$ such that $m/d^3 \gtrsim M/D^3$, the dominant eigenvalue is $\lambda^{(+,1)}=4mG/d^3-MG/D^3$. With the approximation $\lambda^{(+,1)}\approx4mG/d^3$, we find using equation (\ref{vecD}) that
\begin{eqnarray}
\vec{D}\approx
\frac{-2\widetilde{\xi}^{(+,1)}}{M^{\widetilde{r}^{(+,1)}}_{\rm Pl}}\left(\frac{4mG}{d^3}\right)^{(\widetilde{r}^{(+,1)}+1)/2}
\left(\begin{array}{c}1\\ 0 \\0 \end{array} \right),
\end{eqnarray}
where we see that the vector is align with the dominant sources. For an electron at rest (where $\vec{P}=\vec{0}$) we have
\begin{equation}
\mathcal{H}_{NR}=\vec{\sigma}\cdot \vec{D}\approx\frac{-2\widetilde{\xi}^{(+,1)}}{M^{\widetilde{r}^{(+,1)}}_{\rm Pl}}\left(\frac{4mG}{d^3}\right)^{(\widetilde{r}^{(+,1)}+1)/2}\sigma_x.
\end{equation}
The difference of the two eigenvalues of $\mathcal{H}_{NR}$ is then
\begin{equation}
\Delta E\approx\frac{4\widetilde{\xi}^{(+,1)}}{M^{\widetilde{r}^{(+,1)}}_{\rm Pl}}\left(\frac{4mG}{d^3}\right)^{(\widetilde{r}^{(+,1)}+1)/2} .
\end{equation}

\begin{figure}[ptb]
\begin{center}
\includegraphics[width=4.5in]{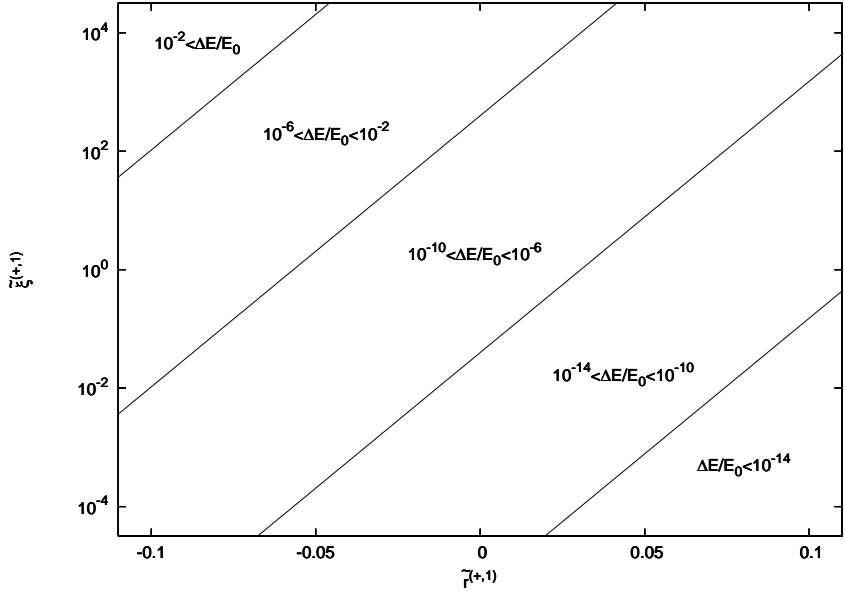}
\caption{Dominant part of the energy difference of the two spin orientations of an electron coupled with the gravitational tidal field generated by the Earth and two $25 \operatorname{kg}$ particles located $10 \operatorname{cm}$ away from the probe according to the model presented in this work ($\Delta E$) over the energy of the electron due to its interaction with the
magnetic field of the Earth ($E_0\approx3\times 10^{-9} \operatorname{eV}$) as a function of
two free parameters of the model: $\widetilde{r}^{(+,1)}$ and $ \widetilde{\xi}^{(+,1)}$.}
\label{figure}
\end{center}
\end{figure}

We now focus in the particular situation where $m=25 \operatorname{kg}$ and $d=10 \operatorname{cm}$. In order to study the range of parameters that would be accessible with an instrument of a given precision, we compute the ratio of the energy differences associated with the effect of interest, to the energy of an electron due to the magnetic field of the Earth ($E_0\approx3\times 10^{-9} \operatorname{eV}$). This is shown in figure \ref{figure} as a function of $ \widetilde{\xi}^{(+,1)}$ and $\widetilde{r}^{(+,1)}$ for the chosen values of $m$ and $d$. The plot has centered in the region around $\widetilde{r}^{(+,1)}=0$ which we believe is both interesting and susceptible to experimental exploration. A much more detailed analysis in conjunction with the experimental colleagues is needed to ascertain the range of parameters that can be explored with the existing technology\footnote{The situation considered here is one where even though the tidal effects are small in the intuitive sense, they might not be small in terms of the accuracy of the extreme sensitivity of the existing apparatuses. To avoid any misunderstandings it should be emphasized that we have carried out no experiment at all, but have provided here a simple description of what we believe is an experiment that can be carried out and that we hope some experimental colleagues will take up.}.

We should also mention that these effects could also be looked for in experiments using ultra-cold neutron sources where it is possible to construct bound states of neutrons in the gravitational field of the Earth \cite{Grenoble} and, in the near future, these experiments might become sensitive to gravitational gradients. 

Regarding extraterrestrial observations, the outlook is grimmer because as we indicated, we need to 
consider polarized matter which is normally very sensitive to magnetic influences so there will be in general many known physical effects that might mimic the signals we are looking for. In this regard the exception might be provided by the neutrinos, which have zero magnetic moment in vacuum 
(care need to be taken to discriminate the so called effective magnetic moments that are thought to be induced when neutrinos travel through a medium \cite{neutrinoMedium}). In particular neutrinos in supernovas seem like the most promising field due to the presence of both, high gravitational tides and high polarization\footnote{We thank professor D.V. Ahluwalia for this remark.}. 

\section{Conclusions}\label{Conclusion}

We have made a concrete and well defined proposal for a novel form of possible phenomenological manifestations of quantum gravity. The proposal involves nontrivial couplings of aspects of the Weyl tensor (characterized by its eigenforms and corresponding eigenvalues) to fermions. In principle, the proposal can be directly tested experimentally. 

At this point it is worth discussing the status of the equivalence principle in the proposal. Let us first note that the strict interpretation of the letter of the principle seems to be violated, as gravity would not be eliminated in a free fall laboratory if terms like the kind we are considering are present. On the other hand, the possibility of realizing a truly inertial frame would still be associated with a freely falling frame in the limit of vanishing curvature of 
space-time. In a classical context one can also achieve an inertial frame in the limit of infinitely small space-time extension of the probes, however, this limit has no counterpart in a quantum mechanical world where the probes could not be arbitrarily localized, a fact that makes it difficult to provide a precise, clear and satisfactory version of the equivalence principle for a quantum world \cite{Geometrical}, although there have been certainly proposals in this regard \cite{Lamrzhal}. In this context it is noteworthy that the effects under consideration arise in the case of particles with spin, for which there are independent indications of a fundamental non-commutativity of the components of the system position associated precisely with the spin of the system and which are thought to reflect an essential limit on its localizability (see Ref. \cite{Chryss}). Thus, if we view the nontrivial curvature-matter couplings arising from a fundamental non-localizability of the quantum probes, we could not argue that the equivalence principle is violated as it pertains only to infinitely localized systems \cite{Geometrical}. In fact, even QED radiative corrections can lead to nontrivial coupling of curvature to photons which can be equally said to violate the letter of the equivalence principle \cite{Talk by Podoc}. The proposal considered here can be though of as an analogous effect arising from the interaction of some fundamental quantum gravity degrees of freedom which are not naturally encoded in the metric formalism, and which reflect the fundamental (and yet unknown) Lorentz Invariant granular structure of space-time.

\section*{Acknowledgments}

\noindent This work was supported
in part by DGAPA-UNAM projects IN108103 and IN119808 and CONACyT 43914-F grant.

\end{document}